\documentclass[aps,preprintnumbers,11pt,amsmath,amssymb,nofootinbib]{revtex4}

\usepackage{amsthm}

\usepackage{listings}
\usepackage{mathrsfs}
\usepackage{hyperref}
\hypersetup{hypertex=true,
            colorlinks=true,
            linkcolor=blue,
            anchorcolor=blue,
            citecolor=blue}

\newcommand{\be}{\begin{equation}}
\newcommand{\ee}{\end{equation}}

\usepackage{color}
\definecolor{purple}{rgb}{0.5,0,0.5}

\def\bea{\begin{eqnarray}}
\def\eea{\end{eqnarray}}
\def\beq{\begin{equation}}
\def\eeq{\end{equation}}

\begin{document}

\title{Constraints on Ho\v{r}ava-Lifshitz gravity from GRB 170817A}
\author{Tao Zhang$^{1,2}$}
\author{Fu-Wen Shu$^{1,2,3}$}
\thanks{shufuwen@ncu.edu.cn}
\author{Qing-Wen Tang$^{1,2}$}
\thanks{qwtang@ncu.edu.cn}
\author{Dong-Hui Du$^{1,2}$}
\affiliation{
$^{1}$Department of Physics, Nanchang University, Nanchang, 330031, China\\
$^{2}$Center for Relativistic Astrophysics and High Energy Physics, Nanchang University, Nanchang 330031, China\\
$^{3}$GCAP-CASPER, Physics Department, Baylor University, Waco, TX 76798-7316, USA }

\begin{abstract}
\baselineskip=0.6 cm
\begin{center}
{\bf Abstract}
\end{center}
In this work we focus on a toy model: (3+1)-dimensional Ho\v{r}ava-Lifshitz gravity coupling with an anisotropic electromagnetic (EM) field which is generated through a Kaluza-Klein reduction of a (4+1)-dimensional Ho\v{r}ava-Lifshitz gravity. This model exhibits a remarkable feature that it has the same velocity for both gravitational and electromagnetic waves. This feature makes it possible to restrict the parameters of the theory from GRB 170817A. In this work we use this feature to discuss possible constraints on the parameter $\beta$ in the theory, by analyzing the possible Lorentz invariance violation effect of the GRB 170817A. This is achieved by analyzing potential time delay of gamma-ray photons in this event. It turns out that it places a stringent constraint on this parameter. In the most ideal case, it gives $|1-\sqrt{\beta}|<(10^{-19}-10^{-18})$.
\end{abstract}

\maketitle

\section{Introduction}

General Relativity (GR) and Quantum Field Theory (QFT) are two cornerstones of modern physics, on which most of our understanding of the universe is currently based. As a fundamental symmetry of nature, Lorentz symmetry plays a crucial role in GR and QFT. However, though the symmetry has been tested with high precision in a variety of settings, there are many motivations to consider its possible violation. One of the main motivations comes from the pursuit for an answer to one of the most fundamental problems of modern physics: how to reconcile GR with QFT. Indeed, although GR and QFT achieve great successes phenomenologically, there are many fundamental questions to be answered. For example, GR loses its prediction as one involves singularities of spacetimes. To overcome these problems, it is generally believed that a full quantum theory of gravity is needed. Many Quantum Gravity (QG) models give hints, more or less, that the Lorentz symmetry should be violated at the scale where the QG effects become relevant, see, for example \cite{AmelinoCamelia:1997gz,AmelinoCamelia:1996pj,Ellis:1999jf,Ellis:2000sx,Ellis:2003sd,Gambini:1998it,Carroll:2001ws} for an incomplete list. This gives an important motivation for testing Lorentz violation (LV) (for review please see \cite{Liberati:2009pf}).

More recent progress on these theoretical efforts, namely, the Ho\v{r}ava-Lifshitz (HL) gravity \cite{Horava:2009uw}, has taken this necessity one step further. Since then, we have a more explicit and more systematic way to Lorentz breaking theories of gravity. In addition, HL theory is power-counting and also perturbatively renormalizable \cite{Barvinsky:2015kil}, at a cost of an anisotropic scaling between the time and spatial coordinates
\begin{equation}
x^{i}\rightarrow bx^{i}, t\rightarrow b^{z}t,
\end{equation}
where $z$ is the dynamical critical exponent. Since this anisotropic scaling clearly breaks boost symmetry, the full diffeomorphism invariance of the theory is broken and is replaced by the diffeomorphism subgroup which preserves the foliation, Diff$_{\mathcal F} (\mathcal M)$. It turns out that Lorentz violations are crucial for the improved UV behavior of this theory \cite{Horava:2009uw}. Although it was found that the original theory suffers from stability and strong coupling problems \cite{Charmousis:2009tc,Padilla:2010ge,Blas:2009yd}, a healthy extension can be achieved \cite{Blas:2009qj,Blas:2010hb} (see \cite{Wang:2017brl} for a review ).

Apart from the theoretical considerations, recent developments on high energy astrophysical observation also demonstrate the need to test LV. Specifically, the apparent absence of the Greisen-Zatsepin-Kuzmin (GZK) cut off \cite{Greisen:1966jv,ZK1969} and the so-called TeV-gamma rays crisis, i.e. the apparent detection of a reduced absorption of TeV gamma rays emitted by AGN \cite{Protheroe:2000hp} directly related to the Lorentz symmetry \cite{Toma:2012xa}.

Recently, on the way of searching for possible LV we have a new method: the GW170817, which the first GW event from a binary neutron star merger as detected by LIGO and Virgo Collaboration \cite{TheLIGOScientific:2017qsa}. It was accompanied by a gamma-ray burst (GRB 170817A) with a time delay $\Delta t = (1.74\pm 0.05)s$ as observed by Fermi-GBM \cite{Monitor:2017mdv}. This observation opens the window of multi-messenger astronomy. Meanwhile, this event provides a strong constraint on the propagation speed of GWs by analyzing the observed time delay between GW and GRB signals. This in turn imposes a strong constraint on modified gravity theories such as \cite{Creminelli:2017sry,Ezquiaga:2017ekz,Baker:2017hug,Oost:2018tcv} (see \cite{Ezquiaga:2018btd} for a review). Concerning HL gravity, it was shown in  \cite{Gumrukcuoglu:2017ijh,Gong:2018vbo} that this puts a stringent constraint on the parameter $\beta$ (see Eq. \eqref{1} below) of the theory\footnote{In addition, there are many available constraints on the parameters of the theory, such as the condition of unitarity and perturbative stability \cite{Blas:2009qj}, the constraints from cosmology \cite{Gumrukcuoglu:2017ijh}, the post-Newtonian (ppN) parameters constraints \cite{Will:2005va,Blas:2010hb,Blas:2011zd,Bonetti:2015oda}, the binary pulsars \cite{Yagi:2013ava}, the black hole \cite{Barausse:2011pu,Barausse:2012ny,Barausse:2012qh,Barausse:2013nwa}. }.

However, the above constraint does not consider the LV of gamma ray itself. A basic fact is that, in spite of tight constraint, the LV in matter sector still cannot be completely excluded. As mentioned before, the absence of the GZK cut off, the TeV-gamma rays crisis in AGN, and the time delay of gamma-ray burst \cite{AmelinoCamelia:1997gz,AmelinoCamelia:1996pj} provides a strong motivation to consider the possible LV of gamma-ray photons. Therefore, a natural question is the following: in addition to the constraint on the parameters of modified theories of gravity from GWs, can we have constraint from GRB data? Or can we give a constraint on these parameters by observing the GRBs which accompany with GWs?

In this paper, we make a preliminary attempt on these issues by investigating the HL gravity. Specifically, we want to investigate whether there exists any additional constraints for the propagation speed of GWs from gamma ray itself. To achieve this, we need to know how the LV of gravity affects the LV of matter. There are some efforts in the past few years in this field \cite{Kostelecky:2003fs,Kostelecky:2010ze,Bailey:2006fd}. As to HL gravity, this refers to the issue: what is the general form of matter Lagrangians consistent with the Diff$_{\mathcal F} (\mathcal M)$ symmetry of HL gravity? There are some scenarios to construct the general form of matter Lagrangians consistent with the reduced symmetry group of HL gravity, such as symmetry consistency \cite{Kimpton:2013zb,Zhu:2011yu,Zhu:2011xe}, spectral action approach \cite{Lopes:2015bra} and dimension reduction approach \cite{Bellorin:2018wst,Restuccia:2019xdo}. In this work, we focus on the last scenario, which is a HL gravity in (4+1) dimensions. The electromagnetic-gravitational coupling in the usual (3+1)-dimensional HL gravity is achieved by considering the (4+1)-dimensional HL  gravity and then performing a Kaluza-Klein(KK) reduction to 3 + 1 dimensions \cite{Bellorin:2018wst,Restuccia:2019xdo}. The propagation speed of the GWs and the GRBs in this model can be obtained by considering the propagation of the gravitational tensor mode and the electromagnetic vector mode, respectively. It turns out that in this model they have the same velocity $\sqrt{\beta}c$, no matter in the Minkowski background or the Friedman-Robertson-Walker (FRW) background. We then restrict this parameter by using the observed data from GRBs. Our results show that we can place a constraint on $\beta$ if we trust the constraint imposed on the LV energy scale of the gamma photons by analyzing their time delay.

In the next section we plan to give a very brief review on the HL gravity in (4+1) dimensions and show how to reduce to (3+1) dimensions  through KK mechanism. We also show how to obtain the velocity both for GWs and electromagnetic wave. In section III we discuss the test for LV of photons from GRBs. We obtain the constraints on the energy scale of the LV of photons for variety of GRB events. In section IV we perform a detailed analysis on the constraints on the parameter of the HL gravity by combining the observational data from GRB 170817A. We give conclusions in the last section.

\section{The Ho\v{r}ava-Lifshitz gravity in (4+1) dimensions}
\subsection{The model}
In this section we review the non-projectable HL gravity in (4+1) dimensions and its Hamiltonian foundation. The action of HL gravity in (4+1)dimension is \cite{Bellorin:2018wst},
\bea\label{1}
S(g_{\mu \nu},N_{\rho},N)=\int dt dx^4 N \sqrt{g}\Big[K_{\mu \nu} K^{\mu \nu}-\lambda K^2+\beta \sideset{^{(4)}}{}{\mathop{R}}
 +\alpha a_{\mu} a^{\mu} +V(g_{\mu \nu},N) \Big].
\eea
\bea
K_{\mu\nu}&=& \frac{1}{2N} \left[ \dot g_{\mu\nu} - \nabla_\mu N_\nu - \nabla_\nu N_\mu \right],
\eea
\bea
K &=& g^{\mu\nu} K_{\mu\nu},
\eea
\bea
a_{\mu} &\equiv& \nabla_{\mu} \ln N.
\eea
where ($K_{\mu \nu} K^{\mu \nu}-\lambda K^2$) is the kinetic term and $\mu$, $\nu$=1, 2, 3, 4. We omit the total coefficient $\frac{M_{pl}}{2}$ because it is a global factor and will be eliminated in the latter analysis. The explicit expression of $V(g_{\mu \nu},N)$ is
\bea\label{higher order}
V(g_{\mu \nu},N)=\sum_{z=2}^4 V_z&=&\frac{1}{{M_*}^2}({\eta_1 \sideset{^{(4)}}{}{\mathop{R^2}}+\eta_2 \sideset{^{(4)}}{}{\mathop{R_{\mu\nu}}}\sideset{^{(4)}}{}{\mathop{R^{\mu\nu}}}}+\eta_3 a_{\mu}\triangle a^{\mu}+\eta_4\sideset{^{(4)}}{}{\mathop{R}} \nabla_\mu a_{\mu}+\cdots)\nonumber\\
&&+\frac{1}{{M_*}^4}(\psi_1\sideset{^{(4)}}{}{\mathop{R_{\mu\nu}}} \triangle\sideset{^{(4)}}{}{\mathop{R^{\mu\nu}}}+\psi_2\nabla_\mu\sideset{^{(4)}}{}{\mathop{R_{\nu k}}}\nabla^\mu\sideset{^{(4)}}{}{\mathop{R^{\nu k}}} +\psi_3a_{\mu}{\triangle}^2 a^{\mu}+\cdots)\nonumber\\
&&+\frac{1}{{M_*}^6}(\omega_1\sideset{^{(4)}}{}{\mathop{R^4}}+\omega_2\sideset{^{(4)}}{}{\mathop{R}}a_\mu\triangle^2a^\mu+\omega_3\sideset{^{(4)}}{}{\mathop{R}}\sideset{^{(4)}}{}{\mathop{R_{\mu\nu}}} \triangle\sideset{^{(4)}}{}{\mathop{R^{\mu\nu}}}+\cdots).
\eea
where $\eta_i$, $\psi_i$ and $\omega_i$ are coupling constants, $\nabla_\mu$ is the covariant derivative with $\Delta=\nabla_\mu\nabla^\mu$. $M_*$ is the HL energy scale, and it has $M_*\gtrsim 10^{10}\sim10^{11}$ GeV \cite{Blas:2010hb}. Even the most energetic particles of the GRB events under consideration is $E \sim 1955$ GeV (GRB 190114C), which is much less than $M_*$. Therefore, the IR terms in \eqref{1} is dominated and $V(g_{\mu \nu},N)$ can be neglected. In what follows, we ignore these terms.

From the Hamiltonian formalism of the Ho\v{r}ava gravity  \cite{Donnelly:2011df,Bellorin:2011ff,Bellorin:2012di,Mukohyama:2015gia,Kluson:2016qrc,Bellorin:2019yvk}, we can obtain the Hamiltonian density $\cal H$ via a Legendre transformation. After adding the primary constraints times Lagrange multipliers it can be written as
\beq\label{2}
\mathcal{H}=\sqrt{g}N \big[\frac{\pi_{\mu\nu}\pi^{\mu\nu}}{g}+\frac{\lambda}{{1-4\lambda}} \frac{\pi^2}{g}-\beta \sideset{^{(4)}}{}{\mathop{R}}-\alpha a_{\mu}a^{\mu}\big]+2\pi^{\mu\nu}\nabla_\mu N_\nu+\sigma P_N.
\eeq
where
\beq
\pi^{\mu\nu}=\partial \mathcal{L} / \partial \dot{g}_{\mu\nu}=\sqrt{g}(K^{\mu\nu}-\lambda g^{\mu\nu}K),
\eeq
is the conjugate momentum to $g_{\mu\nu}$. $\sigma$ and $N_\nu$ are the Lagrange multipliers of the primary constraints (the action (\ref{1}) does not contain the time derivatives $\dot{N}$ and $\dot{N^\rho}$. See \cite{Donnelly:2011df,Bellorin:2011ff} for more details). Variations with respect to other variables generate a complete set of field equations \cite{Bellorin:2018wst}.

The theory has four propagation degrees of freedom: one for the pure gravity and one for pure electromagnetic physical degrees of freedom, and another two for the scalar modes. One of the scalar modes is intrinsic to the HL theory depending on the value of $\lambda$ in the theory ($\lambda=1/4$ is a kinetic conformal point of the theory and this mode is absent at this point). The other mode refers to the dilaton scalar field associated with the KK approach. Since we are interested in the propagation of gravity and electromagnetic modes, and the scalar modes do not influence the propagation of these two modes \cite{Bellorin:2018wst,Restuccia:2019xdo}, in what follows we only focus on the gravity mode and electromagnetic mode.

Following \cite{Bellorin:2018wst,Restuccia:2019xdo}, in order to couple electromagnetism and gravity in a spacetime admitting the same symmetry Diff$_{\mathcal F} (\mathcal M)$, we perform a KK reduction and reduce the theory to (3 + 1) dimensions. We decompose the 4-dimensional Riemannian metric $g_{\mu\nu}$ to a 3-dimensional Riemannian metric $\gamma_{ij}$ and the anti-symmetric electromagnetism vectors $A_i$ \cite{Bellorin:2018wst}
\begin{equation}
g_{\mu\nu}=\left(
  \begin{array}{cc}
    \gamma_{ij}+\phi A_iA_j &  \phi A_j \\
    \phi A_i &  \phi \\
  \end{array}
\right),
\end{equation}
where $\phi$ is a scalar field. After a canonical transformation \cite{Bellorin:2018wst} and taking the form in Ref. \cite{Donnelly:2011df} with the condition $\partial_4=0$, the Hamiltonian density (\ref{2}) becomes
\bea\label{(3+1)dimension}
\mathcal{H}&=&\frac{N}{\sqrt{\gamma \phi}} \big[\phi^2 p^2+p^{ij}p_{ij}+\frac{p^i p_i}{2\phi}+\frac{\lambda}{(1-4\lambda)} \big({p^{ij}\gamma_{ij}+p\phi}\big)^2-\phi\beta\gamma \sideset{^{(4)}}{}{\mathop{R}}-\gamma\phi\alpha a_ia^i\big]\nonumber\\
&&-N_4\mathcal{H}^4 -N_j\mathcal{H}^j-\sigma P_N.
\eea
Here $p_{ij}, p_i$ and $p$ are the conjugate momenta of $\gamma_{ij}, A_i$ and $\phi$ respectively. $N_4$ and $N_j$ are the Lagrange multipliers corresponding to the constraints
\beq
\mathcal{H}^4=\partial_i p^i=0,
\eeq
\beq\label{h constraint}
\mathcal{H}^j=\nabla_i p^{ij}-\frac{1}{2}p^i\gamma^{jk}F_{ik}-\frac{1}{2}p\gamma^{ij}\partial_i\phi=0,
\eeq
and
\beq
\sideset{^{(4)}}{}{\mathop{R}}=\sideset{^{(3)}}{}{\mathop{R}}-\frac{\phi}{4}F^{ij}F_{ij}-\frac{2}{\sqrt{\phi}}\nabla_i \nabla^i \sqrt{\phi},
\eeq
where $\sideset{^{(3)}}{}{\mathop{R}}$ is the curvature of 3-dimensional metric. Variations with respect to variables $\gamma_{ij}$, $A_i$, $\phi$, $p^{ij}$, $p^i$, $p$ gives the equations of motion \cite{Bellorin:2018wst}. The concrete expressions can be found in appendix \ref{appendix1}.

For the Minkowski background we have
\bea\label{background}
\bar{\gamma}_{ij}=\delta_{ij}, \bar{p}_{ij}=0, \bar{N}=1, \bar{A}_i=\bar{p}_i=0, \bar{N}_i=\bar{N}_4=0.
\eea
Under this background ansantz we can get a solution for $\bar{\phi}$ and $\bar{p}$ from the equations of motion. It turns out $\bar{\phi}=1,\ \bar{p}=0$ is the solution in the Minkowski background (For FRW background, however, it has $\bar{\phi}=1$ and $\bar{p}\neq0$ as shown in appendix \ref{appendix2}). Since we are only interested in obtaining a theory where only the gravitational and electromagnetic degrees of freedom propagate and the scalar degree of freedom is non-dynamical, the ground state $\bar{\phi}=1$ and $\bar{p}=0$ is enough. Actually, as we show clearly in the next subsection, even the linear perturbations of $\bar{\phi}$ and $\bar{p}$ are taken into considerations, the wave equations for gravitational and electromagnetic fields are unaffected. Hence this assumption is valid at least to the linear order.

Once we fix the background, i.e., $\bar{\phi}=1$ and $\bar{p}=0$, we find that the parameter $\beta$ (coefficient of $\sideset{^{(4)}}{}{\mathop{R}}$) in (4+1) dimensions (\ref{2}) is compatible with the corresponding parameter ($\beta \bar{\phi}=\beta$ of $\sideset{^{(3)}}{}{\mathop{R}}$) in (3+1) dimensions (\ref{(3+1)dimension}). In fact, except for $\beta$, both $\alpha$ and $\lambda$ are the same for $(4+1)$ dimensions  and its $(3+1)$ KK reduction as can be seen in (\ref{2}) and \eqref{(3+1)dimension}. Therefore, any constraints of these parameters imposed from $(3+1)$ spacetimes are valid also for $(4+1)$, and vice versa.

\subsection{Linear perturbations in Minkowski spacetime}
Now let us make a perturbative analysis on this theory. Perturbations around the Minkowski background (\ref{background}) are given by
\bea\label{perturbs}
\gamma_{ij}=\delta_{ij}+\epsilon h_{ij},  ~ p_{ij}=\epsilon\Omega_{ij},\ N_i=\epsilon n_i,\ N_4=\epsilon n_4,\nonumber\\
N=1+\epsilon n,\ A_i=\epsilon\xi_i,\ p_i=\epsilon\zeta_i,\ \phi=1+\epsilon\tau,\ p=\epsilon\chi,
\eea
where $h_{ij}$, $\Omega_{ij}$, $n_i$, $n_4$, $n$, $\xi_i$, $\zeta_i$, $\tau$, $\chi$ are linear perturbations of each filed. Then the equations of motion at the linear order of $\epsilon$ can be obtained, the details can be found in Ref. \cite{Bellorin:2018wst}.

We decompose the electromagnetic vector and gravitational tensor as follows:
\bea
A_i&=&A_i^T+\partial_i A^L,\label{Ad}\\
h_{ij}&=&h_{ij}^{TT}+\frac{1}{2}\Big(\delta_{ij}-\frac{\partial_i\partial_j}{\Delta}\Big)h^T+\partial_i h_j^T-\partial_j h_i^L.\label{hD}
\eea
Imposing the transverse traceless (TT) gauge conditions $\partial_i A_i^L=0$ and $A^L=0$, combining the equations of motion and the constraint conditions, one finally obtains the wave equations for electromagnetic vectors
\beq\label{3}
\ddot{\xi}_i^T-\beta c^2\Delta\xi_i^T=0.
\eeq

Similarly, imposing the TT gauge $h_{kk}^{TT}=\partial_kh_{ki}^{TT}=0$ and the transverse gauge $h_i^L=0$, one obtains the equations for gravitational waves
\beq\label{4}
\ddot{h}_{ij}^{TT}-\beta c^2\Delta h_{ij}^{TT}=0.
\eeq
These equations (\ref{3}) and (\ref{4}) explicitly show that the speed of gravitational waves and electromagnetic wave are the same  $\sqrt{\beta}c$. The wave equation \eqref{3} implies that EM observations provide an independent approach to constraint the parameter $\beta$, if we consider the time delay of, say, the gamma-ray photons. Eq.\eqref{4} shows, on the other hand, that the GWs provide another way in restricting $\beta$, if we consider the time delay between GWs and GRBs. This makes it very natural to consider GW events with EM counterpart. Under this framework, we achieve this by using data from GW170817 and GRB 170817A.

One more remark here. The scalar perturbation equations can be obtained in the same way. However, it turns out that the wave equations of vector modes and tensor modes \eqref{3} and \eqref{4} are unaffected by the scalar modes. Since we are interested in the propagations of vector modes and tensor modes, we will not write their explicit forms here.

\subsection{Linear perturbations in FRW spacetime}\label{frw perturbation}
In this subsection, we would like to study the perturbation in the FRW background. In appendix \ref{appendix2} we find that the theory with $\lambda=1/3$ admits the following FRW solution
\bea\label{frw background}
\bar{\gamma}_{ij}=a(t)^2\delta_{ij},\  \bar{p}_{ij}=0,\  \bar{p}=-a(t)^2\dot{a}(t),\  \bar{N}=1,\  \bar{A}_i=\bar{p}_i=0,\  \bar{N}_i=\bar{N}_4=0,\  \bar{\phi}=1.
\eea
Perturbations around the FRW background are
\bea\label{perturbs}
\gamma_{ij}&=&a(t)^2(\delta_{ij}+\epsilon h_{ij}),\ p_{ij}=\epsilon\Omega_{ij},\ p=-a(t)^2\dot{a}+\epsilon\chi,\\
N_i&=&\epsilon n_i,\ N_4=\epsilon n_4,\ N=1+\epsilon n,\  A_i=\epsilon\xi_i,\  p_i=\epsilon\zeta_i,\  \phi=1+\epsilon\tau.
\eea
Substituting these perturbations into  Eqs. (\ref{appendix a1})- (\ref{appendix a6})  and expanding them to  the linear order of $\epsilon$, we then get the perturbative equations for vector perturbations
\bea
\dot{\xi}_i&=&\frac{1}{a(t)^3}\zeta_i+\partial_in_4,\\
\dot{\zeta}_i&=&2\frac{\dot{a}(t)}{a(t)}\zeta_i+a(t)\beta\partial_j(\partial_j\xi_i-\partial_i\xi_j),
\eea
and the equations for tensor perturbations
\bea
\dot{h}_{ij}&=&\frac{1}{a(t)}\Big[2\Omega_{ij}-2\delta_{ij}\Omega+2a(t)^4\dot{a}(t)\Big(n+\frac{\tau}{2}-\frac{h}{2}\Big)-2a(t)^2\delta_{ij}\chi\Big]+\frac{1}{2}(\nabla_in_j+\nabla_jn_i),\\
\dot{\Omega}_{ij}&=&\frac{\dot{a}(t)}{a(t)}\delta_{ij}\Omega+\frac{2\dot{a}(t)}{a(t)}\Omega_{ij}+\frac{\beta a(t)^5}{2}\triangle h_{ij}-\frac{\beta a(t)^5}{2}\big(\delta_{ij}-\frac{1}{a(t)^2}\frac{\partial_i\partial_j}{\triangle}\big)\triangle h\nonumber\\
&&-\beta a(t)^5\big(\delta_{ij}-\frac{1}{a(t)^2}\frac{\partial_i\partial_j}{\triangle}\big)\triangle(n+\frac{\tau}{2}),
\eea
where $\triangle=\partial_i\partial_i/a(t)^2$. Taking the transverse traceless (TT) gauge conditions, the wave equations of EM vector and GW tensor can be obtained respectively
\bea
\ddot{\xi}_i^T+H\dot{\xi}_i^T-\beta\triangle\xi_i^T=0,\\
\ddot{h}_{ij}^{TT}+3H\dot{h}_{ij}^{TT}-\beta\triangle h_{ij}^{TT}=0,
\eea
where $H=\dot{a}(t)/a(t)$ is the Hubble parameter. From the wave equations one can show that the speed of gravitational waves and electromagnetic waves in the FRW spacetime are still the same, i.e.,  $\sqrt{\beta}c$, and they are the same as the ones for Minkowski background.

\section{Lorentz violation from gamma-ray burst photons }
In GR, the constancy of the light speed is a basic assumption and the Lorentz invariance keeps well in low energy. However, many quantum gravity theories speculate that the dispersion relation should be modified and the velocity of particles will depend on energy, thus Lorentz invariance will be broken at the approaching Planck scale $(E_{pl}\approx1.22\times10^{19}$ GeV). In this section, we phenomenologically investigate the LV of the GRBs from modified dispersion relation.

For a particle propagating in the quantum spacetime with energy $E\ll E_{pl}$, the dispersion relation can be modified in a general form as a Taylor series
\beq
E^2=p^2c^2+m^2c^4-s_n E^2\Big(\frac{E}{E_{LV,n}}\Big)^n,
\eeq
where $n=1$ or $2$ correspond to linear or quadratic dependence of energy respectively. $s_n =\pm1$ is the sign factor of LV correction and it is determined by experiments \footnote{Most of GRB and AGN observations favor $s_n=1$ except, for example,  \cite{Wei:2017qfz} gives $s_n=-1$. However, the observations of Crab pulsar can have two possibilities ($s_n =\pm1$) \cite{Zitzer:2013gka,Gaug:2017hgh,Ahnen:2017wec}.}. For photon events, it indicates whether the high-energy photons travel slower ($s_n=1$) or faster ($s_n=-1$) than the low-energy photons, $m$ is the rest mass of the particle and $E_{LV,n}$ is the $n$th-order LV energy scale to be determined by experiments.

GRB 170817A is a photon event with vanishing rest mass, then the relation can be written as
\beq\label{dr}
E^2=p^2c^2-s_n E^2\Big(\frac{E}{E_{LV,n}}\Big)^n.
\eeq
Assuming that the relation $\upsilon=\partial E/\partial P$ still hold in quantum gravity, the propagation velocity will be modified
\beq\label{dr}
\upsilon(E)=c\Big[1-s_n\frac{n+1}{2}\Big(\frac{E}{E_{LV,n}}\Big)^n\Big].
\eeq
The vaule $E_{LV,n}$ can be determined through observing the time lag between two photons with different energy \cite{AmelinoCamelia:1997gz}, or the spectral lag that the arrival time delay between light curves with different energy band.
From \eqref{dr}, we get the expression of $E_{LV,n}$
\beq
E_{LV,n}^n=s_n \frac{1+n}{2H_0}\frac{E_h^n-E_1^n}{\Delta t_{LV}}\int_0^z \frac{(1+z^\prime)dz^\prime}{h(z')},
\eeq
where $h(z)=\sqrt{\Omega_\Lambda+\Omega_K(1+z)^2+\Omega_m(1+z)^3+\Omega_R(1+z)^4}$ and $z$, $H_0$, $E_h$ are the redshift, the Hubble constant and the energy of the high-energy photon, respectively. $E_1$ is the energy of the low energy photon, which is generally treated as the energy of the trigger photon. However, compared to the high-energy photons, this energy is too low and cab be ignored in general. $\Delta t_{LV}$, which is the time lag due to the Lorentz violation effect, admits the relation
\beq
\Delta t_{LV}=\Delta t_{obs}-(1+z)\Delta t_{in},\label{tlv}
\eeq
where $\Delta t_{in}$ denotes the emission time lag between $E_h$ and $E_1$ in the source, whose value depends on the intrinsic mechanism and model. Appearance of factor $1+z$ in front of $\Delta t_{in}$ is because $\Delta t_{in}$ is measured in the rest frame of the source, while both $\Delta t_{LV}$ and $\Delta t_{obs}$ are in the observer's frame. It usually assumes that all $\Delta t_{in}$ of GRBs are the same \cite{Ellis:2005wr}.

Rewriting (\ref{tlv}) in the following form is helpful
\beq
\frac{\Delta t_{obs}}{(1+z)}=\frac{K(z)}{E_{LV,n}^n}+\Delta t_{in},
\eeq
where
\beq
K(z)=s_n \frac{1+n}{2H_0}\frac{E_h^n-E_1^n}{(1+z)}\int_0^z \frac{(1+z^\prime)dz^\prime}{h(z')}.
\eeq

In this work we only consider the linear effect ($n=1$), so from (\ref{dr}) we find the LV effect is
\beq\label{5}
|\Delta\upsilon|\equiv \Big|1-\frac{\upsilon(E_{high})}{c}\Big|=\frac{E_{high}}{E_{LV,1}}.
\eeq

The leading-order LV energy scale $E_{LV,1}$ is determined by astronomical observations such as gamma-ray bursts, active galactic nuclei (AGN) and Crab pulsars, etc.. In what follows we only use the data from GRBs such that we can have more consistent result as we are interested in the GRB 170817A. In a series of literatures, a set of constraints have been obtained for different GRB events. We list these results in TABLE \ref{t1} \footnote{In this paper we only consider the $E_{LV,n}$ obtained from time lag analysis, since compared to the spectral lag, it puts much more stringent constraint on the LV effect due to  the lower energy band  for those events of spectral lag \cite{RodriquezMartinez:2006xc,Bolmont:2006kk,Ellis:2005wr,Chang:2015qpa,Wei:2016exb,Bernardini:2017tzu,Wei:2017qfz}.}. They show that the bound ranges from $10^{17}$ GeV to $10^{20}$ GeV.

\begin{table}
\caption{The lower bound on $E_{LV,1}$ from different GRB samples in the literatures. (S: short burst, L: long burst, L,S: both.)}
\begin{center}
\begin{tabular}{|c|c|c|c|}
\hline
   GRB samples & bound on $E_{LV,1}\texttt{(GeV)}$ & $|\Delta\upsilon|=E_{high}/E_{LV,1}$ \\ \hline
   080916C(L) \cite{Tajima:2009az}   & $1.3\times10^{18}$ & $(10^{-17}-10^{-16})$ \\ \hline
   090902B(L),090926A(L),080916C(L) \cite{Shao:2009bv}   & $(2.2\pm0.2)\times10^{17}$ & $(10^{-16}-10^{-15})$ \\ \hline
   090902B(L),090926A(L),080916C(L),090510(S) \cite{Shao:2009bv}   & $(2.2\pm0.9)\times10^{17}$ & $(10^{-16}-10^{-15})$ \\ \hline
   090902B(L)\footnote{In \cite{Chang:2012gq}, the second energetic photon with $E_{high} = 11.16$ GeV
of GRB 090902B was chosen to calculate $E_{LIV,1}$, compared to \cite{Shao:2009bv} where  $E_{high} = 33.4$ GeV.},090926A(L),080916C(L),090510(S) \cite{Chang:2012gq}   & $1\times10^{20}$ & $(10^{-19}-10^{-18})$ \\ \hline
   130427A(L) \cite{Amelino-Camelia:2013naa}   & $(4.6\pm0.5)\times10^{17}$ & $(10^{-16}-10^{-15})$ \\ \hline
   5 GRBs(L) \cite{Zhang:2014wpb}   & $(3.05\pm0.19)\times10^{17}$ & $(10^{-16}-10^{-15})$ \\ \hline
   7 GRBs(L,S) \cite{Zhang:2014wpb}   & $(3.2\pm0.9)\times10^{17}$ & $(10^{-16}-10^{-15})$ \\ \hline
   13 GRBs(L,S) \cite{Zhang:2014wpb}   & $(5.7\pm2.5)\times10^{17}$ & $(10^{-16}-10^{-15})$ \\ \hline
   8 GRBs(L) \cite{Xu:2016zxi}   & $(3.6\pm0.26)\times10^{17}$ & $(10^{-16}-10^{-15})$ \\ \hline
   8 GRBs(L,S) \cite{Ellis:2018lca}   & $(2.4\sim8.4)\times10^{17}$ & $(10^{-16}-10^{-15})$ \\ \hline
   190114C(L) \cite{Acciari:2020kpi}   & $5.8\times10^{18}$ & $(10^{-17}-10^{-16})$ \\ \hline
\end{tabular}\label{t1}
\end{center}
\end{table}

Generally speaking, different analysis methods and energy section or cosmological parameters in $h(z)$ may affect the bound value of $E_{LV,1}$. As an example, as shown in TABLE II, for the same event GRB 090510, different analysis methods or different energy section from light curve lead to different $\Delta t_{LV}$ \cite{Ackermann:2009aa,Ghirlanda:2009mj,Vasileiou:2013vra,Couturier:2013bja,Vasileiou:2015wja}, and different parameters in $h(z)$ \cite{Xiao:2009xe} also give a slightly different results. From TABLE II we see that the influences of these factors are limited and thus we can ignore them in general.
\begin{table}
\caption{The lower bound on $E_{LV,1}$ from GRB 090510 obtained from different methods.}
\begin{center}
\begin{tabular}{|c|c|c|c|}
\hline
  sample &  bound on $E_{LV,1}\texttt{(GeV)}$ &  $|\Delta\upsilon|=E_{high}/E_{LV,1}$ \\ \hline
  090510(S) \cite{Ackermann:2009aa} & $ \text{several}\ 1.22\times10^{19}$ &  $ (10^{-18}-10^{-17})$ \\ \hline
  090510(S) \cite{Ghirlanda:2009mj} & $(1.61\ \text{or}\ 1.3)\times10^{19}$ &  $(10^{-18}-10^{-17})$ \\ \hline
  090510(S) \cite{Xiao:2009xe} & $7.72\times10^{19}$ &  $(10^{-18}-10^{-17})$ \\ \hline
  090510(S) \cite{Vasileiou:2013vra} & $9.3\times10^{19}$ &  $(10^{-18}-10^{-17})$ \\ \hline
  090510(S) \cite{Couturier:2013bja} & $(7.6 \ \text{or}\  2.4)\times10^{19}$ &  $(10^{-18}-10^{-17})$ \\ \hline
  090510(S) \cite{Vasileiou:2015wja} & $3.4\times10^{19}$ &  $(10^{-18}-10^{-17})$ \\ \hline
\end{tabular}
\end{center}
\end{table}

We now turn to the possible LV effect from GRB 170817A's observations by combining the data observed by Fermi-GBM \cite{Monitor:2017mdv}. The accurate way is to get the possible velocity deviation (LV effect) by performing the time or spectral lag analysis. However, the Fermi-GBM data of the GRB 170817A is not significant enough to perform the time or spectral lag analysis \cite{Goldstein:2017mmi}, and we cannot obtain $E_h$ by two ways above. Theoretically, we can estimate the possible LV effect of this event by combining those results as listed in TABLE I. To do this, we first estimate the highest energy by gamma-gamma absorption with the bulk Lorentz factors $\Gamma$ in the early phase, such as 10 seconds before the GBM trigger, the details are presented in \cite{Tang:2014awa}
\beq
E_{high}=\frac{\Gamma^2(m_e c^2)^2}{E_{high,an}(1+z)^2},
\eeq
where $z$ is the redshift of the GRBs, $m_e$ is the mass of electron, $E_{high,an}$ is the spectral break/cutoff energy and $E_{high}$ is the corresponding highest energy of photon.
We assume $E_{high,an}=E_{peak}$ (the peak energy of the Band fit of the GBM data), which implies those photons have the maximum probabilities to undergo the absorption with the highest-energy photons ($E_{high}$). For the Lorentz factor, we employ the estimated values $(100-10^{2.5})$ as suggested in the literatures \cite{Wu:2018bxg,Lamb:2018qfn,Gill:2019tfm,Hajela:2019mjy}, leading to an estimation that the range of the highest energy $E_{high}$ is $(10-200)$ GeV (the observations $E_{peak}=(185\pm62)$ keV and $z\simeq0.001$ in \cite{Monitor:2017mdv}). Based on this estimation and (\ref{5}), we expect that the GRB 170817A suggests a possible LV effect $|\Delta\upsilon|\sim(10^{-19}-10^{-18})$ if the most stringent constraint $E_{LV,1}\sim10^{20}$ GeV \cite{Chang:2012gq} is used. The full LV effects are listed in the TABLES I and II (as it is an estimation value, we only list the magnitude of the effect).

\section{Constraints on parameters from GW \& GRB}

Now we have come to a stage that we can analyze and discuss possible constraints on the parameters of the theory from GRB 170817A. In particular, we are going to impose constraints on the parameter $\beta$ of the HL gravity.

Many constraints, theoretically and experimentally, were placed on the parameters of the (3+1)-dimensional HL theory in the last few years. For example, the unitarity and perturbative stability \cite{Blas:2009qj} require $\frac{3\lambda-1}{\lambda-1}>0$, $\beta>0$ and $0<\alpha<2\beta$. The post-Newtonian (ppN) analysis \cite{Blas:2010hb,Blas:2011zd} shows $\left|4(\alpha+2-2\beta)\right|\lesssim 10^{-4}$ and $\left|\frac{\alpha+2-2\beta}{2\beta-\alpha}\frac{(\alpha-2\beta)(2\lambda-1)+3\lambda-1}{\lambda-1}\right|\lesssim 10^{-7}$. Recent GW experiments, which observed a $\Delta t=(1.74\pm0.05)s$ time delay between GW170817 and GRB 170817A, give a stringent constraint on the speed deviation between GWs and the light (GRBs). After assuming two signals was emitted simultaneously or the GRB signal was emitted 10 seconds after the GW signal, we have \cite{Monitor:2017mdv}
\beq\label{8}
-3\times10^{-15}\leq\frac{\upsilon_{GW}-\upsilon_{GRB}}{\upsilon_{GRB}}\leq 7\times10^{-16}.
\eeq
By assuming that GRBs travels at the speed of light (i.e., no LV for the GRB photons), one gets a constraint on $\beta$ as \cite{Gumrukcuoglu:2017ijh}
\beq\label{13}
|1-\sqrt{\beta}|<10^{-15}.
\eeq
This in turn gives a bound on $\alpha$ as $\left|\alpha\right|\lesssim 10^{-4}$ and $\lambda\sim 1/3$.

In our model, however, $\beta$ is unconstrained (at the linear order) simply from (\ref{8}), since GWs and GRBs travel at the same velocity $\sqrt{\beta}c$. The fact that phenomenological analysis made in the last section shows that the GRB itself is possibly Lorentz invariance violated implies that one may place constraints on $\beta$ by using GRB 170817A. To be more precisely, we notice that
from HL gravity (\ref{1}) one can obtain a modified dispersion relation which is of the form
\beq\label{hldr}
E^2=\beta p^2c^2+\mathcal{O}\big(\frac{p^4c^4}{M_\ast^2}\big),
\eeq
where $M_\ast$ is the HL energy scale as introduced in Eq.(\ref{higher order}). To the leading order we find $\upsilon_{GW}=\upsilon_{GRB}=\sqrt{\beta}c$.  Comparing to the phenomenological result (\ref{dr}), one immediately gets
\beq
\beta=\Big[1-s_n\frac{n+1}{2}\Big(\frac{E}{E_{LV,n}}\Big)^n\Big]^2.
\eeq

As a result, if we apply the most stringent one of the LV effect for GRB 170817A obtained in the last section, i.e. $|\Delta\upsilon| <(10^{-19}-10^{-18})$ to the ($4+1$) HL gravity, we are able to place the constraint on the parameter $\beta$
\beq\label{10}
|1-\sqrt{\beta}|<(10^{-19}-10^{-18}).
\eeq
This is the main result of the present paper.

Several remarks are as follows. Firstly, we should emphasize that this result is only valid for the present HL model which possesses a remarkable feature that the velocity of GWs and GRBs is the same, and restricting the parameter from GRBs is possible. Secondly, due to the absence of the direct observations of the assumed time delay of GRB 170817A, this result severely depends on the value $E_{LV,1}$ obtained from other events and, as shown before, our estimation about $E_{high}$. It has possibility that the new constraint obtained here is overestimated. Thirdly, the bound of $E_{LV,1}$ ranges from $10^{17}$ GeV to $10^{20}$ GeV as listed in TABLE \ref{t1}, which means that we obtain the weaker constraint if we abandon the most stringent bound. In spite of these flaws, this result is very nontrivial, as it shows a possible scenario to consider the constraint from  GRBs, and in principle, this method can be extended to more general models with LV gravity. Moreover, the second and third points can be overcome once we have direct observation of time delay of EM counterpart in the future GW events.

\section{Summary}
In this paper we study (3+1)-dimensional Ho\v{r}ava-Lifshitz gravity coupling with an anisotropic gauge field. This toy model can be generated through a KK reduction of an original (4+1)-dimensional Ho\v{r}ava-Lifshitz gravity. This model exhibits a remarkable feature that it has the same velocity (scales as $\sqrt{\beta}c$) for both gravitational waves and electromagnetic waves. This feature allow us to use the observations of GW's EM counterpart (namely the electromagnetic wave), so as to restrict the parameters of the theory. Specifically, we use this feature to discuss potential constraints on the parameter $\beta$ in the theory, a parameter characterizing how far the theory deviate from the Lorentz symmetry. We achieve this by analyzing the possible LV effect of the GRB 170817A, namely, by analyzing the potential time delay of gamma-ray photons in the GRB 170817A. Our results show that, indeed, the constraint of this parameter is possible as we consider the LV effect of GRBs.

Our present analysis on the LV scale $E_{LV,1}$ only includes GRB observations. In fact, in the past few years there are many astronomical observations such as AGN \cite{Aharonian:2008kz,Biller:1998hg,Albert:2007qk,Martinez:2008ki,Sanchez:2015yua,Biteau:2015qta,Lorentz:2016aiz,Cologna:2016cws} and Crab pulsar \cite{Otte:2012tw,Zitzer:2013gka,Gaug:2017hgh,Ahnen:2017wec} which also put a lower bound on $E_{LV,1}$, for example, from AGN it has $E_{LV,1}>3.3\times10^{19}$ GeV~\cite{Cologna:2016cws} and from Crab pulsar it is $E_{LV,1}>7.8\times10^{17}$ GeV \cite{Ahnen:2017wec}. The most strong constraint till now is presented in a recent paper \cite{Albert:2019nnn}, where the LV energy scale $E_{LV,1}$ is over 1800 times the Planck energy, implies a LV effect $|\Delta\upsilon|\sim(10^{-21}-10^{-20})$ and as a result, $|1-\sqrt{\beta}|<(10^{-21}-10^{-20})$ for our HL model.

\section*{Acknowledgments}

We would like to thank Jorge Bellor\'{i}n for useful communications. This work was supported in part by the National Natural Science Foundation of China under Grant Nos. 11975116, 11903017, 11665016, and Jiangxi Science Foundation for Distinguished Young Scientists under grant number 20192BCB23007.

\appendix
\section{Background solutions of field equations in Minkowski spacetimes}\label{appendix1}

Variation (\ref{(3+1)dimension}) with respect to variables $\gamma_{ij}$, $A_i$, $\phi$ and their conjugate momenta $p^{ij}$, $p^i$, $p$ gives the following six equations of motion \cite{Bellorin:2018wst}
\bea
\dot{\gamma}_{ij}&=&\frac{N}{\sqrt{\gamma\phi}}\big[2p_{ij}+\frac{2\gamma_{ij}\lambda}{1-4\lambda}\big(p^{lm}\gamma_{lm}+p\phi\big)\big]+\nabla_{(i}N_{j)},\label{appendix a1}\\
\dot{A}_i&=&\frac{Np_i}{\sqrt{\gamma\phi^3}}+\partial_iN_4+\frac{1}{2}N_j\gamma^{jk}F_{ik},\label{appendix a2}\\
\dot{\phi}&=&\frac{N}{\sqrt{\gamma\phi}}\big[2p\phi^2+\frac{2\lambda}{1-4\lambda}\big(p^{lm}\gamma_{lm}+p\phi\big)\phi\big]+\frac{1}{2}N^i\partial_i\phi,\label{appendix a3}
\eea
and
\bea
\dot{p}^{ij}&=&\frac{N\gamma_{ij}}{2\sqrt{\gamma\phi}}\big[\phi^2p^2+p^{lk}p_{lk}+\frac{1}{\phi}p^lp_l+\frac{\lambda}{1-4\lambda}\big(p^{lm}\gamma_{lm}+p\phi\big)^2\big]
-\frac{N}{\sqrt{\gamma\phi}}\big[2p^{ik}p^j_k+\frac{1}{2\phi}p^ip^j\nonumber\\
&&+\frac{2\lambda}{1-4\lambda}\big(p^{lm}\gamma_{lm}+p\phi\big)p^{ij}\big]+N\sqrt{\gamma\phi}\beta\big[\frac{R\gamma^{ij}}{2}-R^{ij}\big]
+\beta\sqrt{\gamma}\big[\nabla^i\nabla^j(N\sqrt{\phi})\nonumber\\
&&-\gamma^{ij}\nabla_k\nabla^k(N\sqrt{\phi})\big]
+\frac{\beta N}{2}\sqrt{\gamma\phi^3}\big[F^{in}F^j_n-\frac{\gamma^{ij}}{4}F_{mn}F^{mn}\big]+\beta\sqrt{\gamma}\big[\gamma^{ij}\partial_l N\partial^l\sqrt{\phi}\nonumber\\
&&-2\partial^iN\partial^j\sqrt{\phi}\big]+\alpha N\sqrt{\gamma\phi}\big[\frac{\gamma^{ij}}{2}a_ka^k-a^ia^j\big]-\nabla_k\big[p^{k{(i}}N^{j)}-\frac{p^{ij}N^k}{2}\big]\nonumber\\
&&+\frac{1}{2}N^ip^l\gamma^{jm}F_{lm}+\frac{1}{2}pN^i\partial^j\phi,\label{appendix a4}\\
\dot{p}^i&=&\beta\partial^j\big(N\sqrt{\gamma\phi^3}F^{ji}\big)-\frac{1}{2}\partial_k(N^kp^i-N^ip^k),\label{appendix a5}\\
\dot{p}&=&-\frac{N}{\sqrt{\gamma}}\big[\frac{3}{2}\sqrt{\phi}p^2-\frac{1}{2\sqrt{\phi^3}}p^{ij}p_{ij}-\frac{3}{4\sqrt{\phi^5}}p^ip_i+\frac{\lambda}{1-4\lambda}\big(\frac{3}{2}\sqrt{\phi}p^2+\frac{pp^{ij}\gamma_{ij}}{\sqrt{\phi}}\nonumber\\
&&-\frac{1}{2}\frac{(p^{ij}\gamma_{ij})^2}{\sqrt{\phi^3}}\big)-\beta\gamma\big(\frac{1}{2}\sqrt{\phi}R-\frac{3}{8}\sqrt{\phi}F^{ij}F_{ij}\big)-\frac{\gamma}{2\sqrt{\phi}}\alpha a_ia^i\big]-\beta\sqrt{\frac{\gamma}{\phi}}\nabla_i\nabla^iN\nonumber\\
&&+\frac{1}{2}\partial_i(pN^i).\label{appendix a6}
\eea
The background of the Minkowski space is
\bea\label{a1}
\bar{\gamma_{ij}}=\delta_{ij},\ \bar{p_{ij}}=0,\ \bar{N}=1,\ \bar{N_i}=\bar{N_4}=0,\ \bar{A_i}=\bar{p_i}=0.
\eea
Under this background, we have $\bar{p}=0$ and $\bar{\phi}=constant$. Without loss of the generalization, we set $\bar{\phi}=1$ throughout the paper.

\section{Background solutions of field equations in FRW spacetimes}\label{appendix2}

For the FRW spacetimes, the background metric takes
\bea\label{a2}
\bar{\gamma}_{ij}=a(t)^2\delta_{ij},\ \bar{A}_i=\bar{p}_i=0,\ \bar{N}=1.
\eea
Inserting this background ansatz into (\ref{appendix a1}), (\ref{appendix a2}) and (\ref{appendix a3}), we find that $\bar{N_4}=0$, $\bar{N_i}=0$ and $\bar{\phi}=\bar{\phi}(t)$ is a solution. The explicit form of $\bar{\phi}(t)$ can be obtained by solving (\ref{appendix a4}), (\ref{appendix a5}) and (\ref{appendix a6}).

To proceed, let us make a canonical transformation, then we have
\bea
\bar{p}^{ij}=\frac{\sqrt{\gamma\phi}}{a(t)^2}\Big(\frac{\dot{a}(t)}{a(t)}-\lambda K\Big)\delta^{ij},\ \  \bar{p}_{ij}=a(t)^2\sqrt{\gamma\phi}\Big(\frac{\dot{a}(t)}{a(t)}-\lambda K\Big)\delta_{ij},\ \  \bar{p}=\sqrt{\frac{\gamma}{\phi}}\Big(\frac{\dot{\bar{\phi}}(t)}{\bar{\phi}(t)}-\lambda K\Big),
\eea
where $K=\big(3\dot{a}(t)/a(t)+\dot{\bar{\phi}}(t)/2{\bar{\phi}(t)}\big)$ is the trace of $K_{\mu\nu}$. Substituting these relations into (\ref{appendix a4}), (\ref{appendix a5}) and (\ref{appendix a6}), we can obtain the following solution
\bea
2\dot{a}(t)^2+a(t)\ddot{a}(t)=0,
\eea
and
\bea
\bar{\phi}=\phi_0a(t)^\rho,
\eea
where $\phi_0=\bar{\phi}(t_0)$ and $\rho$ is given by
\bea
\rho&=&\frac{1}{12(\lambda-1)(4\lambda-1)}\big(-8[3+\lambda(16\lambda-13)]+16\times2^\frac{2}{3}(5\lambda-2)(-3+2\lambda(9+2\lambda(5\lambda-8)))/\nonumber\\
&&(\lambda(180+\lambda(207+4\lambda(-1063+2\lambda(1473+500(\lambda-3)\lambda))))+3(-9+\nonumber\\
&&\sqrt{3}\sqrt{-(1-4\lambda)^4(\lambda-1)^2(101+\lambda(-1086+\lambda(4237+1000\lambda(4\lambda-7))))}))^\frac{1}{3}+\nonumber\\
&&8(2\lambda(180+\lambda(207+4\lambda(-1063+2\lambda(1473+500(\lambda-3)\lambda))))+6(-9+\nonumber\\
&&\sqrt{3}\sqrt{-(1-4\lambda)^4(\lambda-1)^2(101+\lambda(-1086+\lambda(4237+1000\lambda(4\lambda-7))))}))^\frac{1}{3}\big).
\eea

The expression of $\rho$ can be extensively simplified for $\lambda=\frac{1}{3}$. In this case we have $\rho=0$ and thus $\phi_0=\bar{\phi}(t_0)$. Without loss of generality, we can set $\bar{\phi}(t_0)=1$. In the present paper, for simplicity,  we only focus on this case. The case with $\lambda\neq\frac{1}{3}$, though complicated, is similar. In summary, we get a background solution in the FRW spacetime which is of the following form
\bea\
\bar{\gamma}_{ij}=a(t)^2\delta_{ij},\  \bar{p}_{ij}=0,\  \bar{p}=-a(t)^2\dot{a}(t),\  \bar{N}=1, \ \bar{A}_i=\bar{p}_i=0,\  \bar{N}_i=\bar{N}_4=0,\  \bar{\phi}=1.
\eea

\end{document}